\documentstyle[12pt,aaspp4,flushrt]{article}
\begin{document}
\parindent=1.0cm

\title{THE NEAR-INFRARED PHOTOMETRIC PROPERTIES OF BRIGHT GIANTS 
IN THE CENTRAL REGIONS OF THE GALACTIC BULGE}

\author{T. J. Davidge \altaffilmark{1}}

\affil{Canadian Gemini Project Office, Herzberg Institute of 
Astrophysics, \\ National Research Council of Canada, 5071 W. Saanich Road, \\ 
Victoria, BC Canada V8X 4M6\\ {\it email:tim.davidge@hia.nrc.ca}}

\and

\affil{Department of Physics \& Astronomy, University of British 
Columbia, \\ Vancouver, BC Canada V6T 1Z4}

\altaffiltext{1}{Visiting Astronomer, Cerro Tololo Inter-American Observatory. 
CTIO is operated by AURA Inc. under contract to the National Science 
Foundation.}

\begin{abstract}

	Images recorded through broad ($J, H, K$), and narrow (CO, and 
$2.2\mu$m continuum) band filters are used to investigate the photometric 
properties of bright ($K \leq 13.5$) stars in a $6 \times 6$ arcmin field 
centered on the SgrA complex. The giant branch ridgelines in the $(K, J-K)$ and 
$(K, H-K)$ color-magnitude diagrams are well matched by the Baade's Window (BW) 
M giant sequence if the mean extinction is A$_K \sim 2.8$ mag. Extinction 
measurements for individual stars are estimated using the M$_K$ versus infrared 
color relations defined by M giants in BW, and the majority of stars 
have A$_K$ between 2.0 and 3.5 mag. The extinction is locally high in 
the SgrA complex, where A$_K \sim 3.1$ mag. Reddening-corrected CO indices, 
CO$_o$, are derived for over 1300 stars with $J, H$, and $K$ brightnesses, and 
over 5300 stars with $H$ and $K$ brightnesses. The distribution of CO$_o$ 
values for stars with $K_o$ between 11.25 and 7.25 can be reproduced using the 
M$_K -$ CO$_o$ relation defined by M giants in BW. The data thus suggest 
that the most metal-rich giants in the central regions of the bulge and 
in BW have similar photometric properties and $2.3\mu$m CO strengths. Hence, 
it appears that the central region of the bulge does not contain a population 
of stars that are significantly more metal-rich than what is seen in BW.

\vspace{0.3cm}

\end{abstract}

\section{INTRODUCTION}

	The stars in the central regions of the Galaxy 
have been the target of numerous photometric and spectroscopic investigations. 
The brightest, best-studied, objects belong to a young population 
(e.g. reviews by Blum, Sellgren, \& DePoy 1996a, and Morris \& Serabyn 1996) 
that dominates the near-infrared light within 0.1 -- 0.2 
parsec ($\sim 2.5 - 5.0$ arcsec) of SgrA* (Saha, Bicknell, \& McGregor 1996), 
and is concentrated within the central $\sim 1$ parsec (Allen 1994). However, 
there is also a large population of older stars near the Galactic Center (GC) 
belonging to the inner bulge, and it is only recently that efforts have been 
made to probe the nature of these objects.

	Minniti {\it et al.} (1995) reviewed metallicity estimates for a number 
of bulge fields, and a least squares fit to these data indicates that 
$\Delta$[Fe/H]/$\Delta$log(r) $= -1.5 \pm 0.4$ (Davidge 1997). The presence of 
a metallicity gradient suggests that the material from which bulge stars formed 
experienced dissipation, so it might be anticipated that the central 
regions of the bulge will contain the most metal-rich stars in the Galaxy. 
While the fields considered by Minniti {\it et al.} are at relatively 
large distances from the GC, and hence do not monitor trends at small radii, 
observations of other galaxies indicate that 
population gradients can extend into the central regions of 
bulges. This is clearly evident in spectroscopic studies of M31 
(e.g. Davidge 1997), a galaxy that shares some morphological 
similarities with the Milky-Way (e.g. Blanco \& Terndrup 1990).

	It is not clear from the existing observational data if the 
bright stellar contents in the inner bulge and 
Baade's Window (BW), a field that is dominated by stars roughly 0.5 kpc from 
the GC, are similar. The $K$ luminosity function 
(LF) of moderately faint stars within an arcmin of SgrA* has a power-law 
exponent similar to that seen in BW (Blum {\it et al.} 1996a; Davidge {\it et 
al.} 1997). However, the significance of this result 
is low, as the LFs of bright giant branch stars are 
insensitive to metallicity (e.g. Bergbusch \& VandenBerg 
1992). The brightest inner bulge stars, which are 
evolving on the asymptotic giant branch (AGB), have $2\mu$m spectroscopic 
properties reminiscent of bright giants in BW, although detailed measurements 
reveal that for a given equivalent width of near-infrared Na and Ca absorption, 
the inner bulge stars have deeper CO bands than giants in BW (Blum, Sellgren, 
\& DePoy 1996b). While suggestive of differences in chemical 
composition, it should be recalled that these objects are the brightest, most 
highly evolved members of the bulge, and Na can be affected by mixing (e.g. 
Kraft 1994). Hence, the spectroscopic properties of 
these bright red giants may not be representative of fainter objects. 

	As the strongest features in the near-infrared spectra of cool evolved 
stars, the $2.3\mu$m first-overtone CO bands provide an important means of 
probing the stellar content of the inner bulge. Unfortunately, the crowded 
nature of the inner bulge, coupled with the low multiplex advantage offered by 
the current generation of cryogenically-cooled spectrographs, most of which use 
a single long slit, makes a $2\mu$m spectroscopic survey of a large sample of 
moderately faint objects a difficult task at present. Narrow-band imaging, 
using filters such as those described by Frogel {\it et al.} (1978), provides 
a highly efficient alternate means of measuring the strength of CO absorption 
in a large number of objects. In the current paper $J, H, K$, CO and $2.2\mu$m 
continuum measurements of moderately faint ($K \leq 13.5$) stars are 
used to measure the strength of CO absorption in stars within 3 arcmin 
of the GC. To the best of our knowledge, this is the largest survey of stellar 
content in the central regions of the bulge conducted to date. The observations 
and reduction techniques are described in \S 2, while the photometric 
measurements are discussed in \S 3. In \S 4 the line-of-sight extinction to 
these sources is estimated by assuming that they follow the same 
M$_K -$ color relations as M giants in BW. Reddening-corrected CO indices 
are derived, and the distribution of CO indices is compared with that 
predicted if stars in the inner bulge and BW follow similar M$_K -$ CO 
relations. A summary and brief discussion of the results follows in \S 5.

\section{OBSERVATIONS AND REDUCTIONS}

	Two different datasets were obtained for this study. The primary 
dataset was recorded at the CTIO 1.5 metre telescope, while a supplementary 
dataset, covering a smaller field but extending to slightly larger distances 
from SgrA* with deeper photometric coverage, was recorded at the MDM 2.4 
metre telescope.

\subsection{CTIO Data}

	The CTIO data were recorded during the nights of UT 
July 21 and 23 1996 with the CIRIM camera, which was mounted 
at the Cassegrain focus of the 1.5 metre telescope. 
The detector in CIRIM is a $256 \times 256$ Hg:Cd:Te array, 
and optics were installed so that each pixel subtends 0.6 arcsec on a side. 
Images of nine overlapping fields, which together cover a $6 \times 6$ arcmin 
area, were recorded through $J, H, K,$ CO, and $2.2\mu$m continuum filters. 
Four exposures of each field were recorded per filter, with the telescope 
pointing offset in a $5 \times 5$ arcsec square pattern (`dithering') between 
integrations. The total exposure times per field are 80 sec ($4 \times 20$ 
sec) in $J$, 40 sec ($4 \times 10$ sec) in $H$, and 20 sec ($4 \times 5$ sec) 
in $K$ and the narrow-band filters. A number of standard stars from 
the lists published by Casali \& Hawarden (1992) and Elias 
{\it et al.} (1982) were also observed during the 5 night observing 
run. The image quality varied from 1.2 to 1.5 arcsec FWHM.

	The data were reduced using procedures similar 
to those described by Hodapp, Rayner, \& Irwin (1992). A median 
dark frame was subtracted from each image, and the result was divided by 
a dome flat. The dome flats were constructed by subtracting images of 
the dome white spot taken with the flat-field lamps off from those recorded 
with the lamps on. This differencing technique removes thermal artifacts that 
can introduce significant background structure in the images at 
longer wavelengths. A DC sky level was measured and subtracted from each 
flat-fielded image, and the results were normalized to a 1 sec integration 
time before being median-combined on a filter-by-filter basis. The 
offsets introduced at the telescope, coupled with the large number of fields 
observed, cause the median-combination process to filter out stellar 
images while retaining thermal signatures. The median-combined 
thermal structure frames produced in this manner were subtracted 
from the sky-subtracted exposures. The results for each field were aligned 
on a filter-by-filter basis to correct for the offsets introduced by dithering,
and then combined by taking the median value at each pixel position.

	The images of the individual fields were then used to construct a 
mosaic covering $6 \times 6$ arcmin. The FWHM of stars in overlapping portions 
of adjacent images were measured to assess differences in image quality, 
which were removed by convolving those frames having 
the smaller FWHM with a gaussian. Overlapping portions of adjacent frames 
were trimmed. The resulting $K$ mosaic, which has an image 
quality of $\sim 1.5$ arcsec FWHM, is shown in Figure 1. 
Variations in the line-of-sight extinction are plainly evident, 
illustrating the role that differential reddening plays in defining the 
photometric properties of objects near the GC.

\subsection{MDM Data}

	The second dataset was recorded 
with the Ohio State University MOSAIC camera, which was mounted 
at the f/7.5 Cassegrain focus of the 2.4 metre MDM telescope, 
during the night of UT June 20 1997. MOSAIC contains a $1024 \times 
1024$ Alladin InSb array; however, only two of the detector quadrants 
were active, so that the total light-sensitive 
region was $512 \times 1024$ pixels. Each pixel subtended 0.3 
arcsec on a side, so that the imaged field is $2.6 \times 5.1$ arcmin.

	A single field, with SgrA* positioned near one end, 
was observed through $J, H,$ and $K$ filters. A complete observing 
sequence consisted of four exposures per filter, with the telescope 
pointing offset in a $10 \times 10$ arcsec square dither pattern between 
individual integrations. Two sequences were recorded -- one, intended to study 
bright stars, had a total integration time of 12 sec ($4 \times 3$ 
sec) per filter. A second, deeper, sequence was obtained with 
total integration times of 240 sec ($4 \times 60$ sec) in 
$J$, 60 sec ($4 \times 15$ sec) in $H$, and 40 sec ($4 \times 10$ sec) in $K$. 
A number of standard stars from the list published by Casali \& Hawarden 
(1992) were also observed during the 4 night observing run.

	The data were reduced using the procedures 
described in \S 2.1. The final $K$ image, for which the image quality is
$\sim 1.0$ arcsec FWHM, is shown in Figure 2. The upper two-thirds 
of this field overlap with the region observed at CTIO.

\section{PHOTOMETRIC MEASUREMENTS}

\subsection{Stellar brightnesses and LFs}

	The brightnesses of individual standard stars were measured with the
PHOT routine in DAOPHOT (Stetson 1987). The coefficients $\alpha$, $\beta$, 
and $\gamma$ in transformation equations of the form:

\hspace*{4.0cm}$M = m + \alpha X + \beta C + \gamma$

\noindent{were} determined using the method of least squares. 
$M$ and $m$ in this equation are the standard and instrumental brightnesses, 
$X$ is the airmass, and $C$ is the instrumental color. 
The night-to-night scatter in the standard 
star measurements was negligible. The residuals in the 
standard star measurements made at CTIO 
are $\pm 0.02$ mag for all filters. The residuals for the 
standard star measurements made at MDM are $\pm 0.02 (J), \pm 0.06 (H)$, and 
$\pm 0.05$ mag $(K)$.

	The standard stars were selected so as 
to reduce possible sources of systematic error. In particular, 
standard stars were observed over the same airmass range 
as the GC -- this is especially important for the measurements recorded at 
the MDM, as the GC never rises above 2 airmasses at that site. 
In addition, an effort was made to observe standard stars covering 
the broadest possible color range, although none of the standard stars have 
colors as red as the heavily obscured stars near the GC. The calibration of 
the CO index was checked by observing globular clusters in which individual 
stars have published CO measurements (Frogel, Persson, \& Cohen 1983 
and references therein).
  
	Stellar brightnesses were measured with the 
PSF-fitting routine ALLSTAR (Stetson \& Harris 1988), which is part of the 
DAOPHOT (Stetson 1987) photometry package. The calibrated brightnesses and 
colors are in good agreement with published values. 
This is demonstrated in Table 1, where the $K$ brightnesses of four bright 
($K \leq 10$) IRS sources are compared with measurements made by 
Blum {\it et al.} (1996a). The mean difference is $\Delta K = 
-0.07 \pm 0.14$, where the uncertainty is the error in the mean. Comparisons 
at other wavelengths show similar agreement.

	Uncertainties in the photometric measurements were estimated by 
comparing the brightnesses of stars common to both the MDM and CTIO datasets. 
These comparisons indicate that the $K$ measurements 
have an uncertainty of $\pm 0.05$ mag when $K 
= 10.5$, and $\pm 0.21$ mag at $K = 13$. The mean difference 
in brightness between the MDM and CTIO measurements at moderately faint 
brightnesses is $\Delta K = -0.01 \pm 
0.02$ mag, where the difference is in the sense MDM -- CTIO.

	The $J, H,$ and $K$ LFs constructed from the CTIO and MDM datasets are 
plotted in Figure 3. Stars brighter than $K \sim 10$ are saturated in the 
deep MDM exposures, so the bright end of the MDM LFs 
were defined using the short exposure data. However, 
the detector in MOSAIC becomes significantly non-linear for 
count rates in excess of 10000 ADU/sec (Frogel 1998 -- private communication), 
so caution should be exercised when interpreting the bright end ($K \leq 10$) 
of the MDM LFs. The linearity of the detector in CIRIM is well-calibrated, 
and the photometry from the CTIO data is reliable to $K \sim 8$. 

	Incompleteness causes the LFs to depart from power-laws 
at the faint end. For the CTIO data this occurs when 
$J = 17.5$, $H = 15.0$, and $K = 13.5$. Incompleteness becomes 
significant about one mag fainter for the MDM data, owing 
to the longer integration times, better image quality, and larger telescope 
aperture. The completeness limits defined in this manner are averages over the 
entire field, and incompleteness sets in at brighter values in the high density 
SgrA environment. 

	Stars brighter than $K \sim 10$ in Figure 3 belong to the young SgrA 
complex or are disk objects. Indeed, the brightest M giants in BW, which are 
evolving on the AGB, have $K_0 \sim 7$ (Frogel \& Whitford 1987) which, if A$_K 
\sim 3$ mag (DePoy \& Sharpe 1991), corresponds to $K \sim 10$ near the GC. 
For comparison, stellar evolution models predict that the 
red giant branch tip for solar and higher metallicity old populations 
occurs near M$_K \sim -7$ (Bertelli {\it et al.} 1994), 
which corresponds to $K \sim 10.5$ near the GC.

	The dashed lines in Figure 3 show the trend defined by M giants in BW, 
as measured from Figure 17 of Tiede, Frogel, \& Whitford (1995), shifted along 
the vertical axis to match the GC $K$ LFs near the faint end. Blum {\it et al.} 
(1996a) found that the power-law exponent at the faint end of the GC $K$ LF is 
similar to that in BW, and the excellent agreement between the GC and BW 
sequences in Figure 3 is consistent with this finding.

\subsection{Color-Magnitude diagrams}

	The $(K, J-K)$, and $(K, H-K)$ CMDs constructed from the CTIO and deep 
MDM data are shown in Figures 4 and 5. Stars with $K \leq 10$ are saturated 
in the MDM data. The CMDs are dominated by a red plume comprised mainly of 
bulge giants, although a significant number of evolved young stars, which also 
have red colors, are concentrated in the SgrA complex (e.g. Davidge {\it et 
al.} 1997). Objects with $(J-K) \leq 2$ likely belong to the foreground disk.

	The large line-of-site extinction in $J$ (A$_J \sim 7.4$ mag) causes 
sample incompleteness to become significant at the color of the red plume on 
the $(K, J-K)$ CMD when $K \leq 12$. Nevertheless, although retricted to 
relatively bright stars, the $(K, J-K)$ CMD still provides an important means 
of comparing the photometric properties of the brightest giants in the inner 
bulge with those in BW. In the upper panel of Figure 6 the ridgeline defined 
from the CTIO ($K \leq 11.5$) and MDM ($K \geq 11.5$) $(K, J-K)$ CMDs, 
calculated using $\pm 0.25$ mag bins in $K$, is compared with the 
BW M giant sequence listed in Table 3B of Frogel \& Whitford (1987). The 
two sequences can be brought into excellent agreement if 
$E(J-K) = 4.4$ mag towards the inner bulge, which corresponds to A$_K = 
2.9$ mag using the Rieke \& Lebofsky (1985) reddening curve.

	The ridgeline of the red plume on the $(K, H-K)$ CMDs, 
where incompleteness does not become significant until $K \sim 14$, 
is also well matched by the BW M giant sequence. This is demonstrated 
in the lower panel of Figure 6 where the locus of the ($K, H-K$) 
CMD, computed using $\pm 0.25$ mag bins in $K$, 
is compared with the BW M giant sequence from Table 3B of Frogel \& Whitford 
(1987). The two sequences can be brought into excellent agreement if 
$E(H-K) = 1.6$ mag for the inner bulge, which is consistent with the 
$(J-K)$ color excess derived from the $(K, J-K)$ CMD.

\section{EXTINCTION-CORRECTED CO MEASUREMENTS}

	The comparisons in Figure 6 indicate that the brightest red giants in 
the inner bulge and BW have similar photometric properties, suggesting that the 
extinction towards the former can be estimated by adopting the intrinsic colors 
of the latter. Blum {\it et al.} (1996a) assumed that stars near the GC have 
colors that are the same as a `typical' M giant in BW: $(J-H)_0 = 0.7$ 
and $(H-K)_0 = 0.3$. However, the current data samples stars spanning 
a wide range of intrinsic brightnesses, and hence colors, and this 
should be taken into account when computing A$_K$. Therefore, for the current 
study $(J-H)_0$ and $(H-K)_0$ were assigned using the M$_K -$ color relations 
from Table 3B of Frogel \& Whitford (1987). M$_K$ was computed for each star by 
assuming that $\mu_0 = 14.5$ (Reid 1993) and A$_K = 2.9$ mag (\S 3.2). 
$(J-H)_o$ and $(H-K)_o$, and hence E$(J-H)$ and E$(H-K)$, were 
then calculated and used to compute an improved A$_K$ 
based on the Rieke \& Lebofsky (1985) reddening curve. Zero reddening was 
assigned to stars when A$_K \leq 0$. These calculations were done only for 
the stars in the CTIO dataset.

	It is likely that stars near the GC will have a range 
of metallicities, so that a single M$_K -$ color relation will not 
apply to all objects. Given this possibility, it is of interest to 
investigate (1) how the A$_K$ distribution changes if fiducial sequences 
other than those defined by BW giants are adopted, and (2) 
how these changes affect, for example, the distribution of 
reddening-corrected CO indices. Therefore, a second set of A$_K$ values were 
computed using relations between M$_K$ and near-infrared colors 
derived from the Frogel, Persson, \& Cohen (1981) aperture 
measurements of 47 Tuc giants. 

	The histogram distribution of A$_K$ values derived for 
stars in the CTIO field, shown in the upper panel of 
Figure 7, peaks near A$_K \sim 2.8$, which corresponds to A$_V = 25.5$, and 
$E(B-V) = 8.2$. The A$_K$ distribution computed using the 47 Tuc M$_K -$ color 
relations is shown as a dotted line in the top panel of Figure 7. Bright giants 
in 47 Tuc are bluer than stars in BW with the same M$_K$, so the A$_K$ 
distribution computed from the 47 Tuc relations is shifted to slightly 
higher values; however, as demonstrated below, this difference has a negligible 
effect on the distribution of reddening-corrected CO indices.

	A$_K$ varies across the field, and there is a tendency for the 
extinction to be highest in the vicinity of the SgrA complex. The radial 
behaviour of A$_K$ is shown in Figure 8, and it is evident that the 
radially-averaged A$_K$ values, shown in the second 
panel from the top, increase when $r \leq 50$ arcsec. 
The SgrA complex contains a number of hot massive evolved stars, and it 
might be anticipated that these objects will skew $\overline{A_K}$ near SgrA* 
to smaller values. However, the evolved young stars near the GC have excess 
infrared emission, which causes their $H-K$ colors to be similar to 
those of red giants (Davidge {\it et al.} 1997). The mean value of A$_K 
\sim 3.1$ mag derived here for the SgrA complex is in good agreement with 
other reddening estimates, which have relied on smaller 
sample sizes (e.g. DePoy \& Sharpe 1991). That this 
area of locally heavy extinction is coincident with the SgrA complex suggests 
that the excess obscuring material may be local to the GC region, rather 
than occuring in the intervening disk. Nevertheless, the possibility that this 
excess extinction may be due to a fortuitously positioned dust cloud 
anywhere along the sight line can not be discounted. 
If the mean extinction towards the inner bulge is A$_K \sim 2.8$ 
mag, then the excess extinction towards SgrA is $3.1 - 2.8 = 0.3$ mag in $K$.

	Reddening-corrected CO indices, CO$_o$, were 
then computed for stars in the CTIO dataset assuming 
that $\frac{E_{CO}}{E_{BV}} = -0.04$ (Elias, Frogel, \& Humphreys 1985). 
The resulting CO$_o$ distribution, shown in the upper panel of Figure 9, peaks 
near CO$_o \sim 0.28$ and is noticeably asymmetric, with an extended tail 
towards smaller CO$_o$ values. The dotted line in the top panel shows 
the distribution predicted if the 47 Tuc brightness-color relations 
are used to estimate extinction, and the result is very similar to that 
produced with the BW brightness-color relations.

	To investigate the effect that sample incompleteness has on the CO$_o$ 
distribution, a second CO$_o$ distribution was computed for brightnesses 
where the sample is complete. The extinction-corrected $K$ 
LF, shown in the top panel of Figure 10, indicates that the data 
are complete when $K_o \leq 9$, which corresponds to $K \leq 12$. The 
CO$_o$ distribution for stars with $K_o < 9$ is shown in the middle panel 
of Figure 9. The distribution derived from stars with $K_o < 9$ 
is skewed towards higher CO$_o$ values than that 
derived from the entire dataset, as expected given the 
surface-gravity sensitivity of the CO bands. Nevertheless, while a 
Kolmogoroff-Smirnoff test indicates that the two distributions are 
significantly different, both peak near CO$_o = 0.3$.

	The LF in Figure 10 can be used to predict the distribution of CO$_o$ 
values that would result if M giants in BW and near the GC have the same 
CO -- M$_K$ relation. The predicted CO$_o$ distribution for stars with 
$K_o$ between 11.25 and 7.25 (the range of brightnesses covered by M giants 
in BW), calculated using the LF in Figure 10 and the CO--M$_K$ relation from 
Table 3B of Frogel \& Whitford (1987), is shown in the upper panel of Figure 
11. The predicted distribution, which does not include the effects of 
observational errors, is visibly asymmetric, in the same sense as the observed 
distribution. It should also be noted that the predicted CO$_o$ distribution 
automatically includes the effects of sample incompleteness, as the observed 
LF is used to set the number of stars with a given CO value.

	In order to compare the observed and 
predicted CO$_o$ distributions it is necessary to include the effects of 
observational errors in the latter. This was done by convolving the 
predicted CO$_o$ distribution with a gaussian, the width of which was selected 
so that the number of stars with predicted CO indices between 0.16 and 0.28 
matched those observed, and the result is shown in Figure 12. It is evident 
that the observed and predicted CO$_o$ distributions have similar shapes. 
Although not plotted on Figure 12, the peak of the CO$_o$ distribution 
predicted if stars near the GC followed the M$_K -$ CO$_o$ relation 
defined by bright giants in 47 Tuc occurs near CO$_o \sim 0.15$, which is 
much smaller than what is observed. Consequently, the comparison in Figure 12 
indicates that M giants in the central regions of the Galaxy and in BW
follow the same M$_K$ -- CO relation. Moreover, the number of stars 
with CO indices similar to red giants in 47 Tuc must be very small. 

	Sellgren {\it et al.} (1990) and Haller {\it et al.} (1996) found that 
$2.3\mu$m CO absorption in unresolved sources 
weakens within 8 arcsec of SgrA*, a result that has been attributed to the 
presence of young main sequence stars (Eckart {\it et al.} 
1995). The present data can be used to study radial trends among bright 
resolved sources in the SgrA complex, as well as search for radial 
CO$_o$ variations among resolved sources over larger angular scales. 
In Figure 13 the CO$_o$ indices for individual stars are 
plotted as a function of distance from SgrA*. There is significant scatter, 
and to help interpret these results the mean CO$_o$ and $K_o$ values in 20 
arcsec intervals are plotted in the middle and lower panel of this figure, 
with the errorbars showing the error of the mean. $<K_o>$ increases with radius 
when $r < 100$ arcsec, due to the radial decrease in stellar density, and the 
corresponding shift of the completeness limit to fainter values. 
The solid squares in the middle panel show the CO index 
that is appropriate for $<K_o>$, based on the M$_K -$ CO relation for M giants 
in BW.

	It is evident from the middle panel of Figure 13 that $<CO_o>$ drops 
when $r < 20$ arcsec, due to the concentration of bright young blue stars, 
which have little or no CO absorption, in the SgrA complex. However, 
there is excellent agreement between observed and predicted CO$_o$ values for 
distances between 70 and 200 arcsec from SgrA*. While there is a 
tendency for $<CO_o>$ to increase when $r < 220$ arcsec, the significance is 
low, as these measurements are based on only a small number of points. 
Therefore, the SgrA complex aside, there is no evidence for a radial departure 
from the BW M$_K -$ CO relation in this field.

\section{SUMMARY AND DISCUSSION}

	Moderately deep near-infrared images have been used to probe the bright 
($K \leq 13.5$) stellar content in the central $6 \times 6$ arcmin field of the 
Galaxy. The ridgeline of the bulge giant branch on the $(K, J-K)$ and $(K, 
H-K)$ CMDs is well matched by the BW M giant sequence, reddened 
according to the Rieke \& Lebofsky (1985) extinction curve. This 
similarity in photometric properties suggests that the extinctions 
to individual stars in the inner bulge can be estimated by adopting 
the M$_K -$ color relations defined by M giants in BW. The mean extinction 
outside of the SgrA complex, where A$_K = 3.1$ mag, is 
A$_K = 2.8$ mag. The extinction estimates for individual 
stars have been used to generate reddening-corrected CO indices, and
the histogram distribution of CO$_o$ values can be reproduced using the 
M$_K -$ CO$_o$ relation defined by M giants in BW. Therefore, M giants near 
the GC and in BW have similar $2.3\mu$m CO strengths.

	A potential source of systematic error in the procedure 
used to compute A$_K$ is that a single set of M$_K -$ color relations 
have been used, and no attempt has been made to allow for a dispersion in the 
metallicities of giants in the central regions of the bulge. There is a 
selection effect for magnitude-limited samples, in the sense that the 
brightest stars in a field containing an old composite population 
are likely the most metal-rich, so the current data likely do not 
sample the full range of metallicities near the GC. In any event, the 
CO$_o$ distribution does not change substantially when M$_K -$ color relations 
defined by red giants in 47 Tuc ([Fe/H] $\sim -0.7$) are used to estimate 
extinction, indicating that the main conclusions of this paper are 
insensitive to the adopted intrinsic colors of GC giants.

	Another source of systematic error 
is that the A$_K$ values derived in \S 4 require measurements 
in $J, H$, and $K$. This broad wavelength coverage introduces a 
bias against heavily reddened objects, which are faint, and hence 
may not be detected, in $J$. In fact, there are regions where the 
extinction is so high that stars are not detected in $K$. The tendency to 
miss the most heavily reddened stars skews the A$_K$ distribution to 
lower values. One way to reduce, but not entirely remove, this bias is to 
estimate extinction using only $H-K$ colors, for which the number of 
objects is over $3 \times$ greater than those with $J$ measurements. 
Following the procedure described earlier, $(H-K)_0$ was assigned 
to each star using the $M_K - (H-K)$ relation for BW M giants, and the results 
are shown in the lower panels of Figures 7 -- 12, while the 
radial distribution of CO$_o$ values is plotted in Figure 14. 
It is evident from the lower panel of Figure 8 that 
a number of sources with relatively high A$_K$ are added to the sample 
when measurements in $J$ are not required. Nevertheless, 
the impact on the CO$_o$ distribution is negligible, indicating that 
stars with higher than average obscuration near the GC have $2.3\mu$m CO 
strengths that are similar to less heavily reddened stars.

	It should be emphasized that the $2.3\mu$m CO bands provide only one 
diagnostic of chemical composition. Indeed, studies of the 
radial behaviour of the $2.3\mu$m CO bands in the bulges of other 
galaxies reveal weak or non-existant gradients (e.g. Frogel {\it et al.} 
1978), even though many of these systems show line strength gradients at 
optical wavelengths. When interpreting this ostensibly contradictory result 
it should be recalled that the CO measurements are dominated by the brightest 
red stars which, in old populations, will be those that are the most 
metal-rich. If the most metal-rich population in the 
bulges of other galaxies follows a single M$_K -$ CO relation with 
no radial dependence, as appears to be the case in the inner regions of 
the Galactic bulge, then this would help to explain why the 
wide-aperture CO measurements of other systems do not show gradients.

	There are indications that the abundances of some species in the 
spectra of metal-rich giants may vary with distance from the GC. In 
particular, the infrared spectra obtained by Blum {\it et al.} 
(1996b) indicate that $2.3\mu$m CO absorption in GC giants is {\it stronger} 
than in BW stars having the same $2\mu$m Na and Ca absorption line strengths. 
Therefore, given that the CO line strengths 
in giants near the GC are similar to those in BW, then the Blum et al. 
measurements are suggestive of radial changes in [Na/Fe] and 
[Ca/Fe] among bulge stars.

	While this paper has concentrated on bulge stars, a modest disk 
population is also evident in the CMDs, and these data can be used to estimate 
the contribution disk objects make to the near-infrared light output near the 
GC. Objects with $(J-K) \leq 2.5$ and brightnesses between $K \sim 7.5$ (the 
approximate saturation limit of the CTIO data) and $K \sim 12$ 
(the faintest point at which the CTIO $K, J-K$ CMD is complete over a broad 
range of colors) account for 2.6\% of the total light from resolved 
sources within 3 arcmin of the GC. The relative contribution made 
by disk stars would be much lower if dust did not obscure 
the bulge. Assuming that (1) the disk stars are not heavily reddened, and 
(2) stars near the GC are obscured by A$_K \sim 2.8$ mag then, after correcting 
for this extinction, the contribution from disk stars drops to only 0.05\% when 
$K_0 \leq 8.5$.

\vspace{0.3cm}
	Sincere thanks are extended to the referee, Jay Frogel, for providing 
comments that greatly improved the paper.

%\placetable{tbl-1}
%\placetable{tbl-2}

\clearpage

\begin{table*}
\begin{center}
\begin{tabular}{cccr}
\tableline\tableline
IRS \# & K$_{TD}$ & K$_{BSD}$ & $\Delta$ \\
\tableline
7 & 6.44 & 6.55 & $-0.11$ \\
9 & 8.80 & 8.57 & 0.23 \\
16NE & 8.56 & 9.01 & $-0.45$ \\
28 & 9.41 & 9,36 & 0.05 \\
\tableline
\end{tabular}
\end{center}
\caption{Comparison with $K$ measurements obtained by Blum {\it et al.} (1996a)}
\end{table*}

\begin{figure}
\plotone{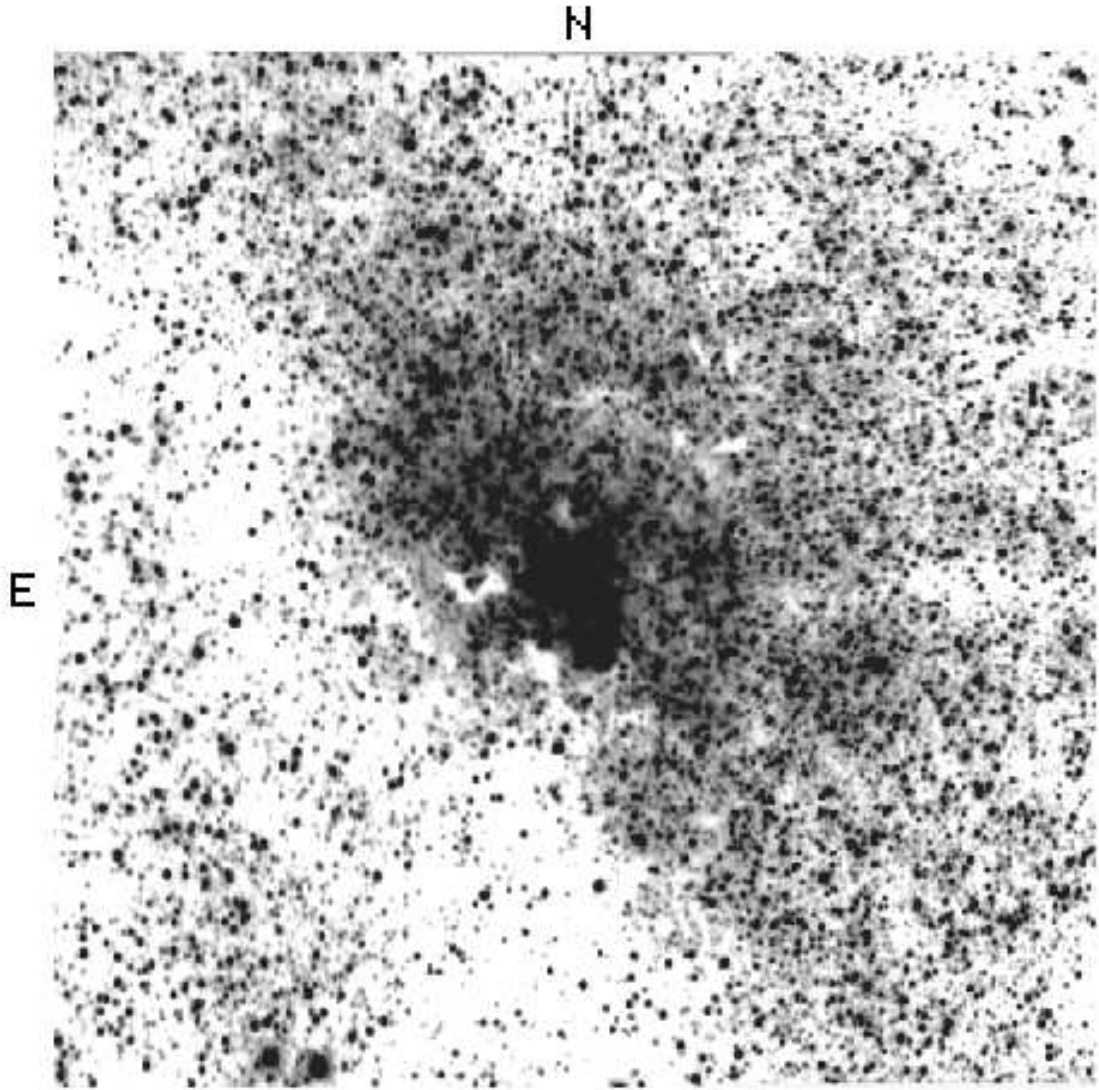}
\caption{The final $K$ image of the $6.17 \times 6.15$ arcmin mosaic 
field obtained at CTIO. The SgrA complex lies near the center of 
the field. The image quality is roughly 1.5 arcsec FWHM.}
\end{figure}

\begin{figure}
\plotone{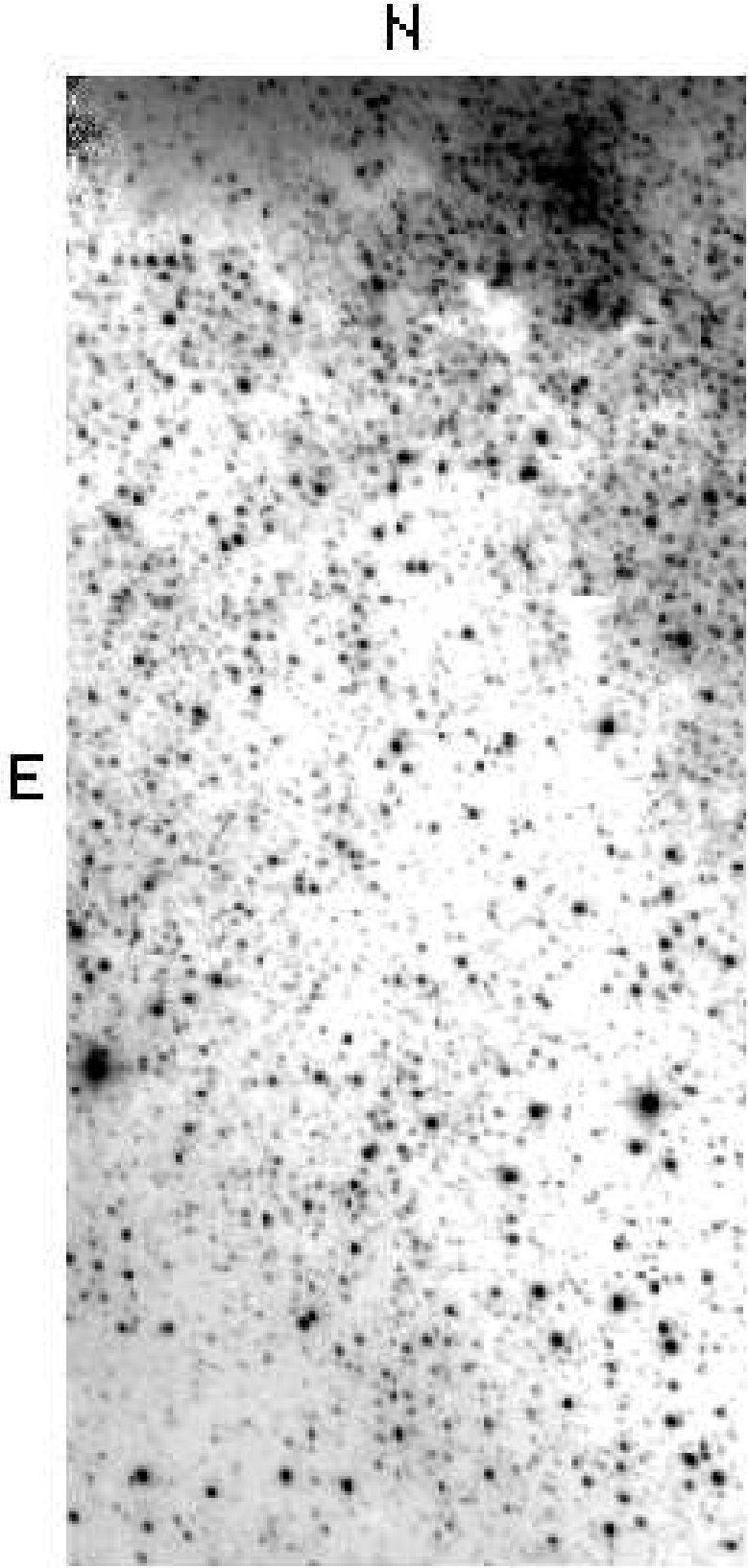} 
\caption{The final $K$ image of the $2.26 \times 4.94$ arcmin field recorded 
with the MDM 2.4 metre telescope. The SgrA complex can be seen 
in the upper northwest corner of the field. The image quality 
is roughly 1.0 arcsec FWHM.}
\end{figure}

\begin{figure}
\plotone{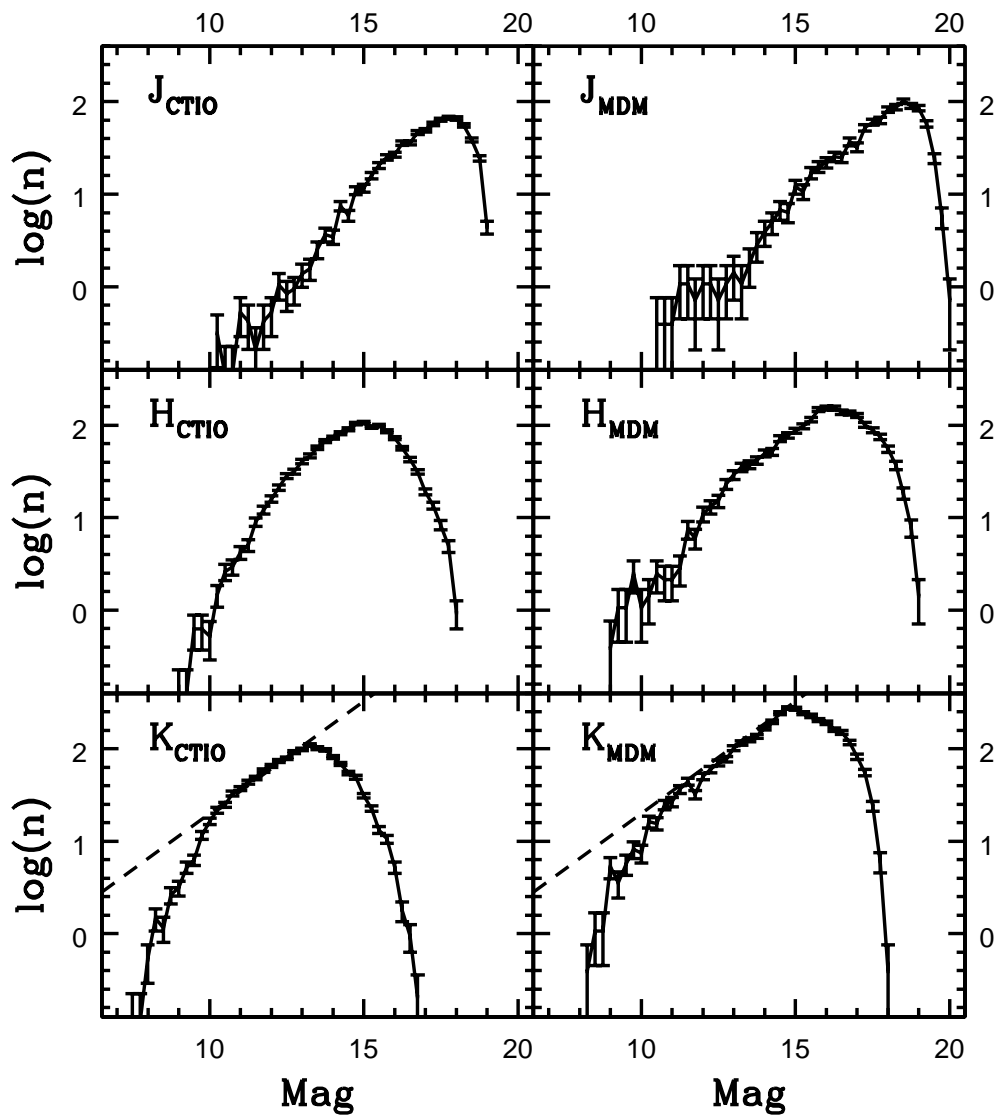} 
\caption{$J, H,$ and $K$ LFs constructed from the CTIO (left column) and 
MDM (right column) datasets. The error bars show the uncertainties due 
to counting statistics. $n$ is the number of stars per square arcmin per 
mag. The dashed line in the bottom row shows the trend defined by bright giants 
in BW, shifted by an arbitrary amount along the vertical axis to match the 
observations between $K = 10$ and 13.}
\end{figure}

\begin{figure}
\plotone{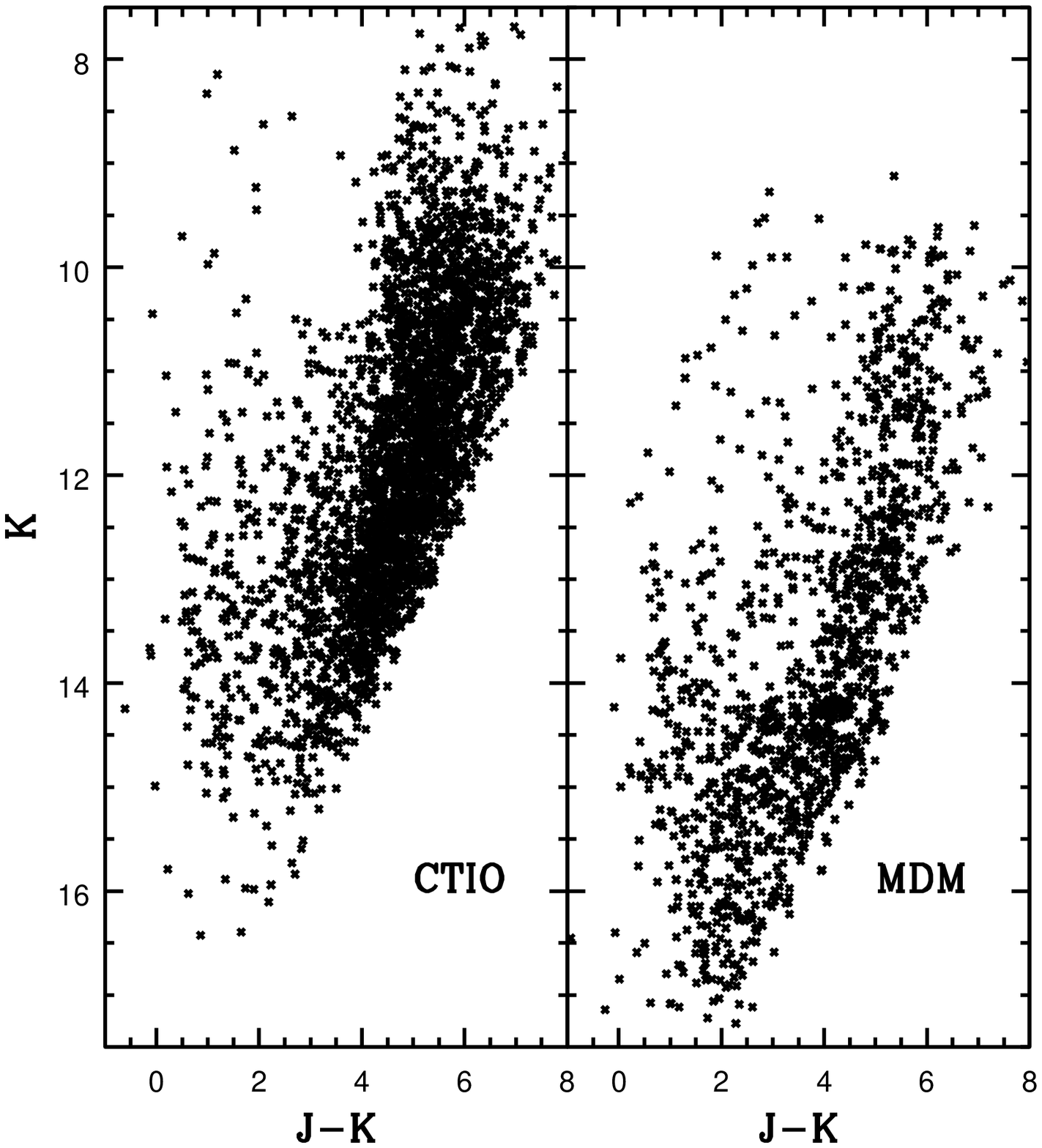} 
\caption{$(K, J-K)$ CMDs for the CTIO and deep (60 sec integration times in 
$J$ and 10 sec in $K$) MDM datasets. Stars brighter than $K = 10$ are 
saturated in the MDM data.}
\end{figure}

\begin{figure}
\plotone{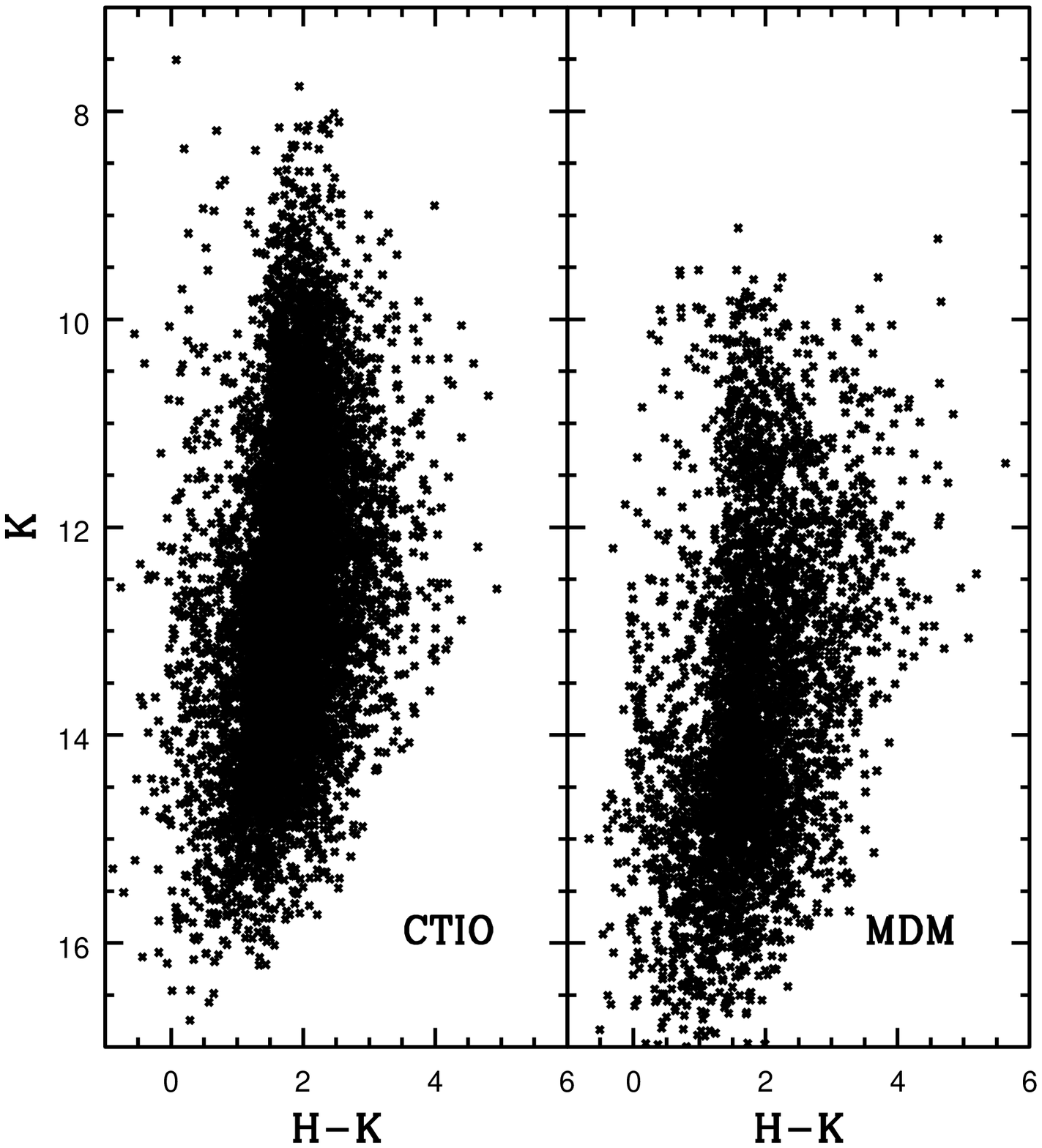} 
\caption{$(K, H-K)$ CMDs for the CTIO and deep (15 sec integration times in 
$H$ and 10 sec in $K$) MDM datasets. Stars brighter than $K = 10$ are 
saturated in the MDM data.}
\end{figure}

\begin{figure}
\plotone{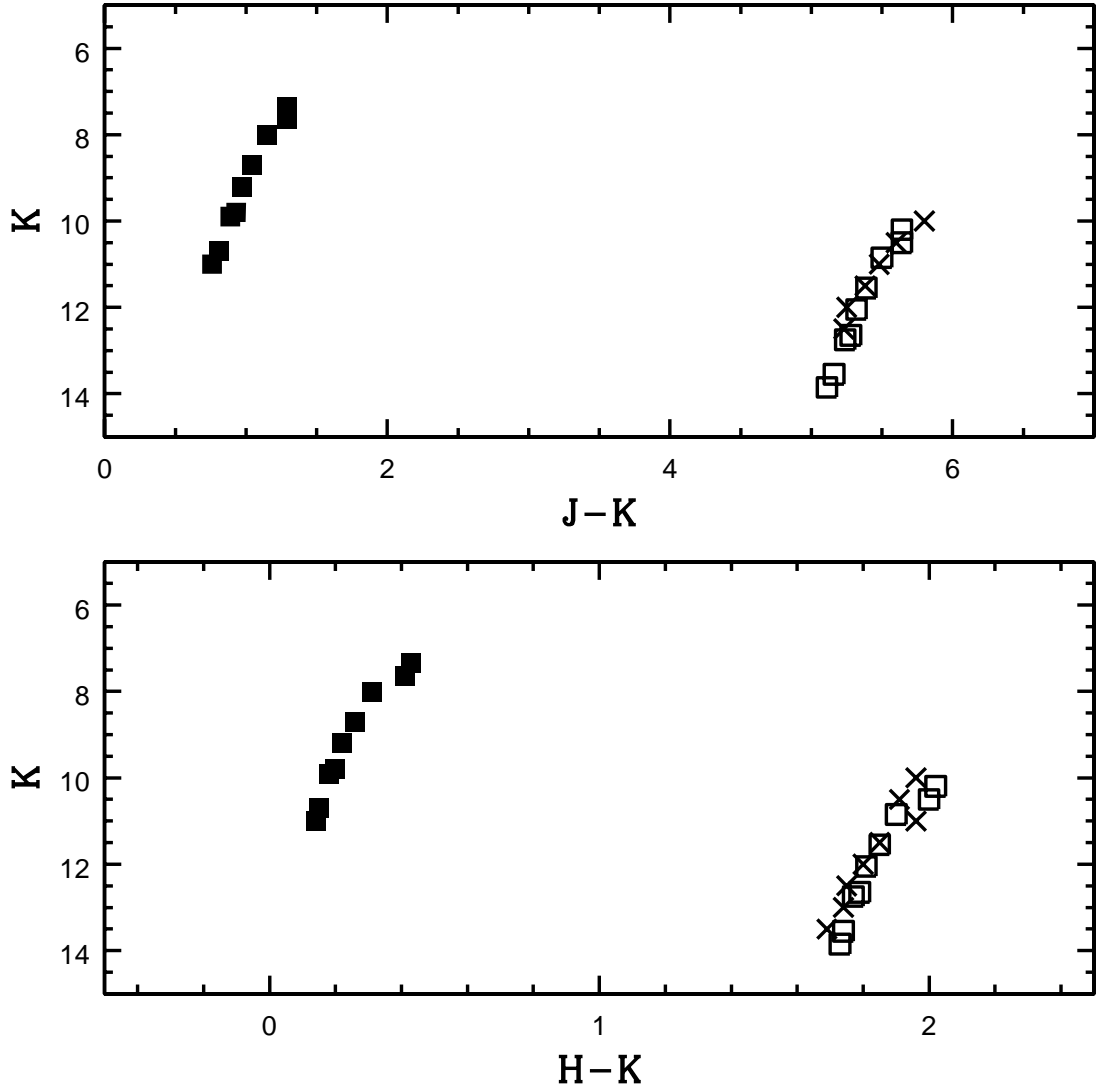} 
\caption{Comparison between the BW M giant sequence 
and the GC giant branch locus on the $(K, J-K)$ (top panel) and 
$(K, H-K)$ (bottom panel) CMDs. The ridgeline of the GC giant branch is shown 
as crosses, while the filled squares are the BW M 
giant sequence listed in Table 3B of Frogel \& Whitford (1987). The open 
squares show the BW sequence as it would appear if A$_K = 2.9$ mag. Note 
that the GC and BW sequences have very similar shapes.}
\end{figure}

\begin{figure}
\plotone{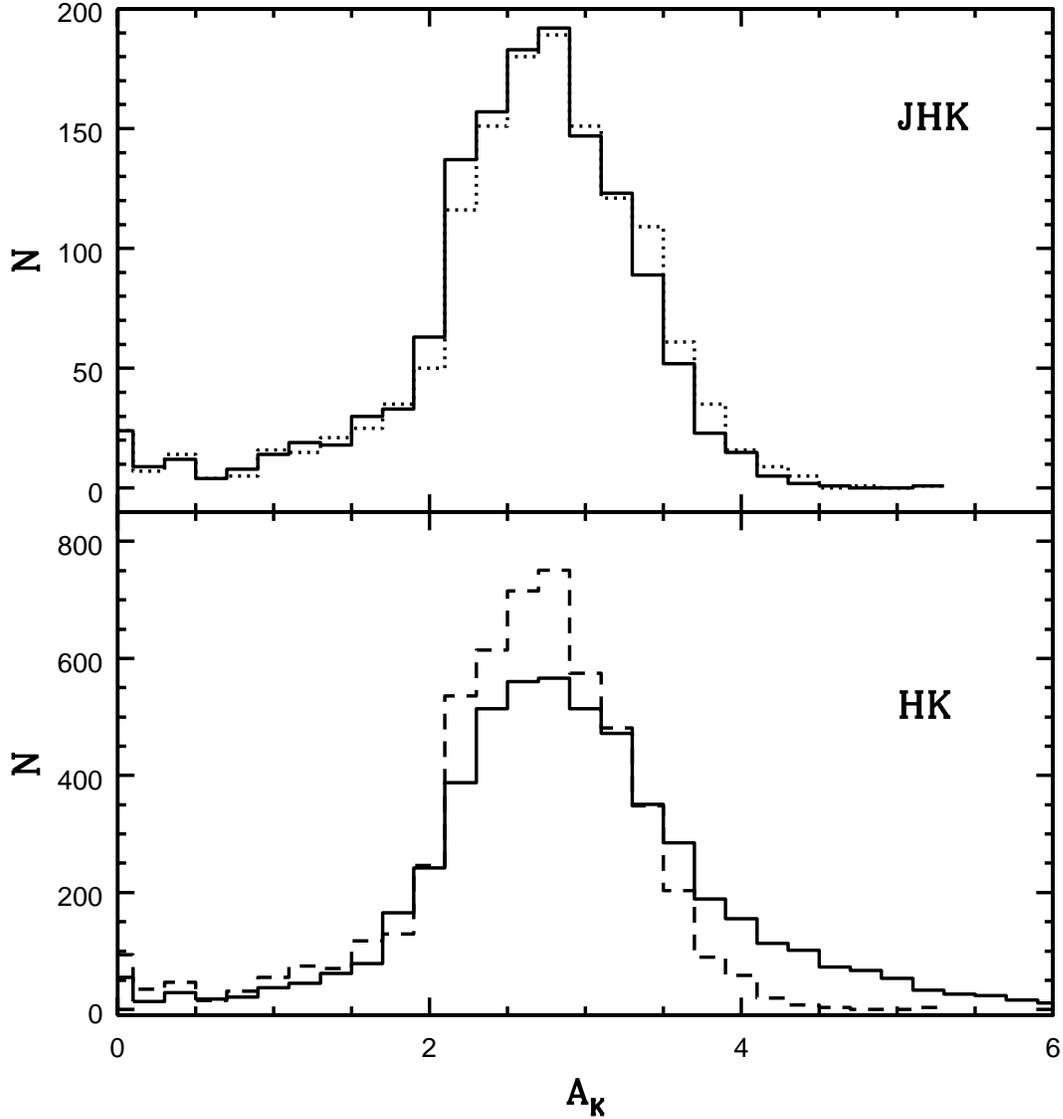} 
\caption{The histogram distribution of A$_K$ values, derived assuming that 
red giants in BW and the inner bulge follow the same M$_K -$ color relations. 
The top panel shows the distribution that results 
if extinctions are computed using both $J-H$ and $H-K$ colors, while the 
lower panel shows the A$_K$ distribution if only $H-K$ colors are 
used. The dotted line in the upper panel is the A$_K$ distribution 
if extinctions are estimated using M$_K -$ color relations defined by 47 Tuc 
giants. The dashed line in the lower panel shows the A$_K$ distribution from 
the upper panel, scaled to match the number of stars having only $H-K$ colors. 
Note that the A$_K$ distribution derived only from $H-K$ includes more highly 
reddened objects, due to the deeper nature of the $H$ and $K$ images.}
\end{figure}

\begin{figure}
\plotone{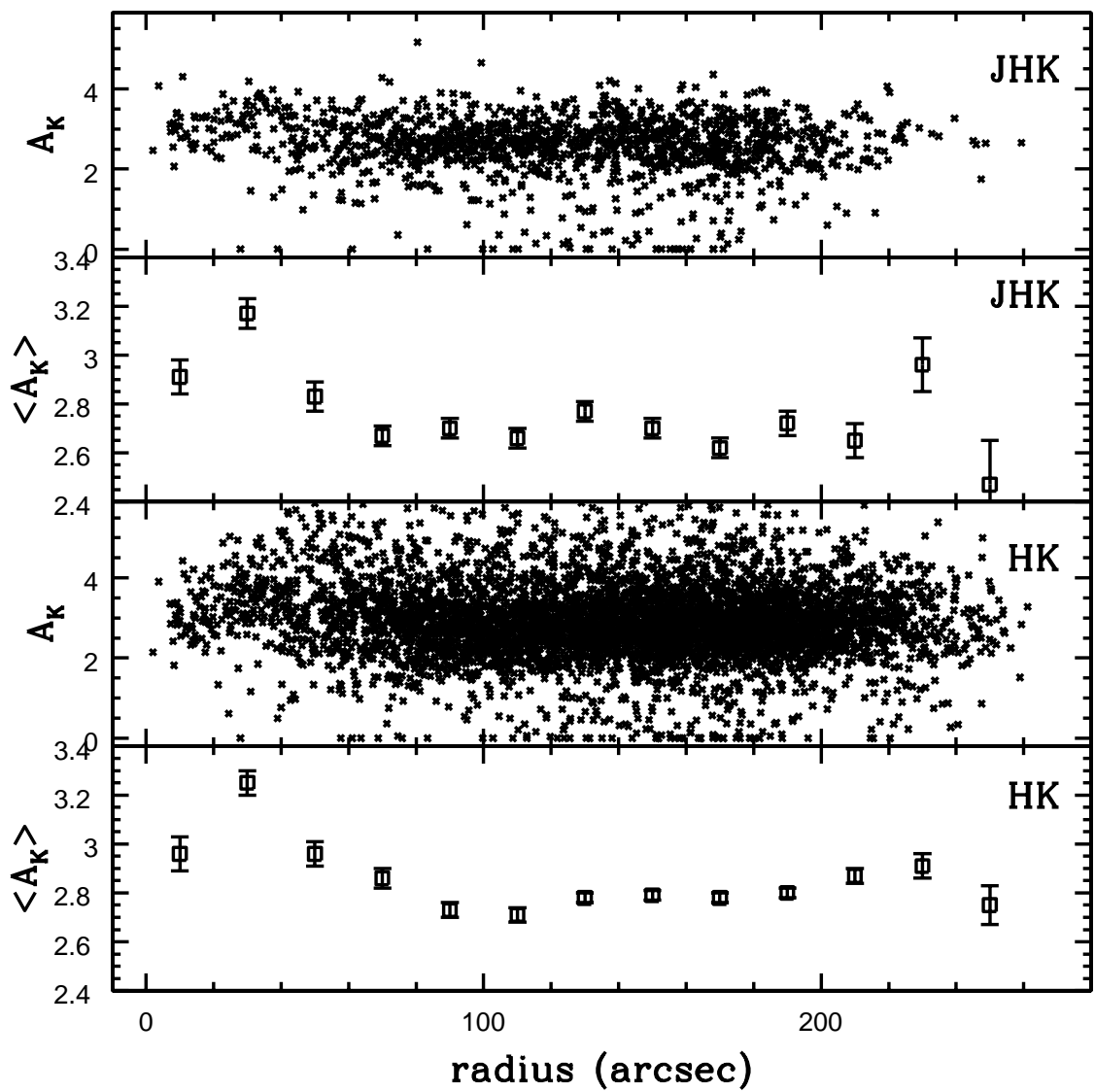} 
\caption{A$_K$ as a function of distance from SgrA*. The top panel shows A$_K$ 
for individual stars computed using both $J-H$ and $H-K$ colors, 
while the means in 20 arcsec annuli, with errorbars indicating the standard 
deviation of the mean, are plotted in the second panel from the top. Note the 
tendency for A$_K$ to increase towards smaller radii. The lower two 
panels show the A$_K$ values that result if only $H-K$ colors are used.}
\end{figure}

\begin{figure}
\plotone{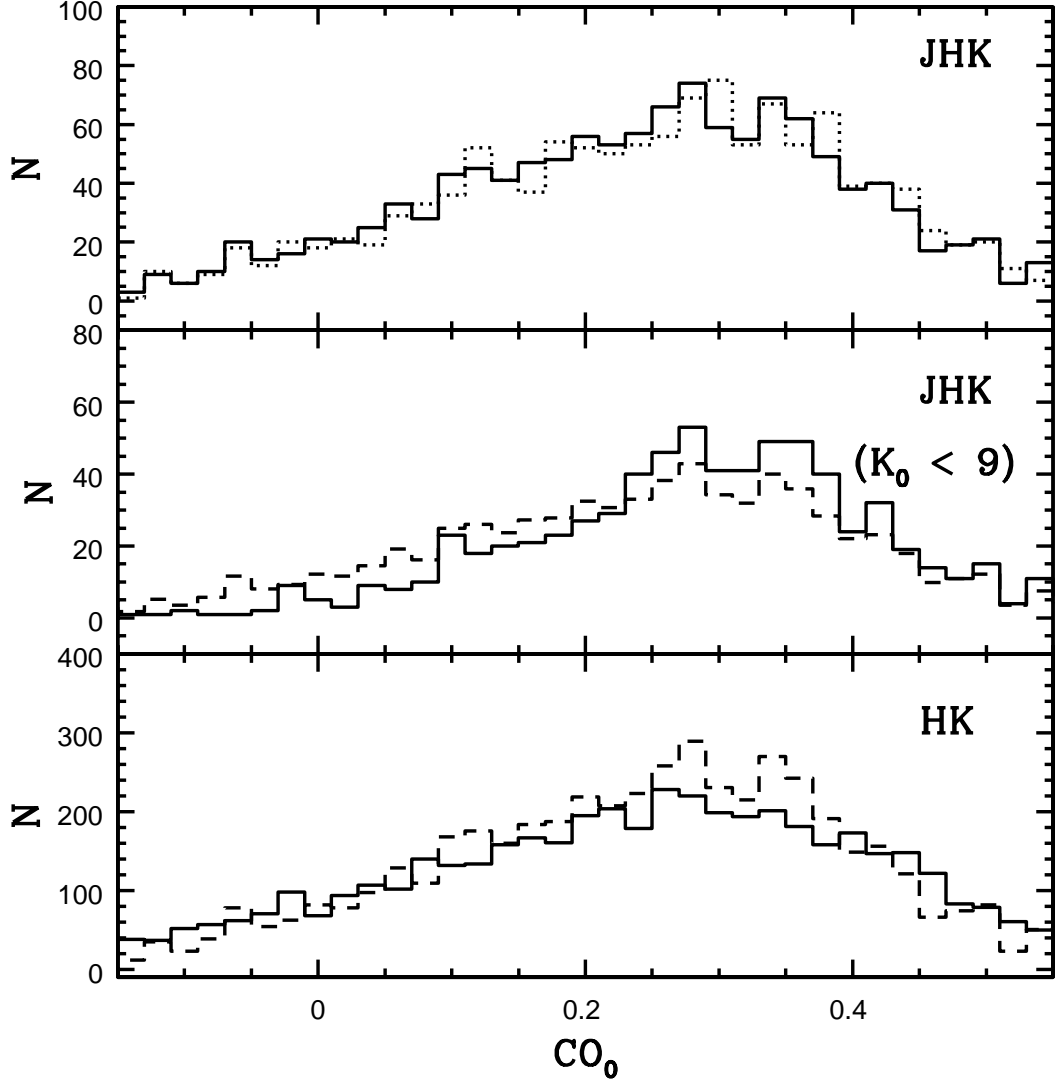}
\caption{CO$_o$ distributions calculated using the 
A$_K$ values shown in Figure 7. The top two panels show the distributions 
that result if extinctions are computed using both $J-H$ and $H-K$ colors, 
while the lower panel shows the CO$_o$ distribution if only 
$H-K$ colors are used. The distribution in the middle 
panel is restricted to stars with $K_o < 9$, where the data are complete. 
The dotted line in the upper panel is the CO$_o$ distribution that results 
if extinction is estimated from the 47 Tuc M$_K -$ color relation. The dashed 
lines in the lower two panels show the CO$_o$ distribution from the upper 
panel, scaled to correct for differences in sample size.}
\end{figure}

\begin{figure}
\plotone{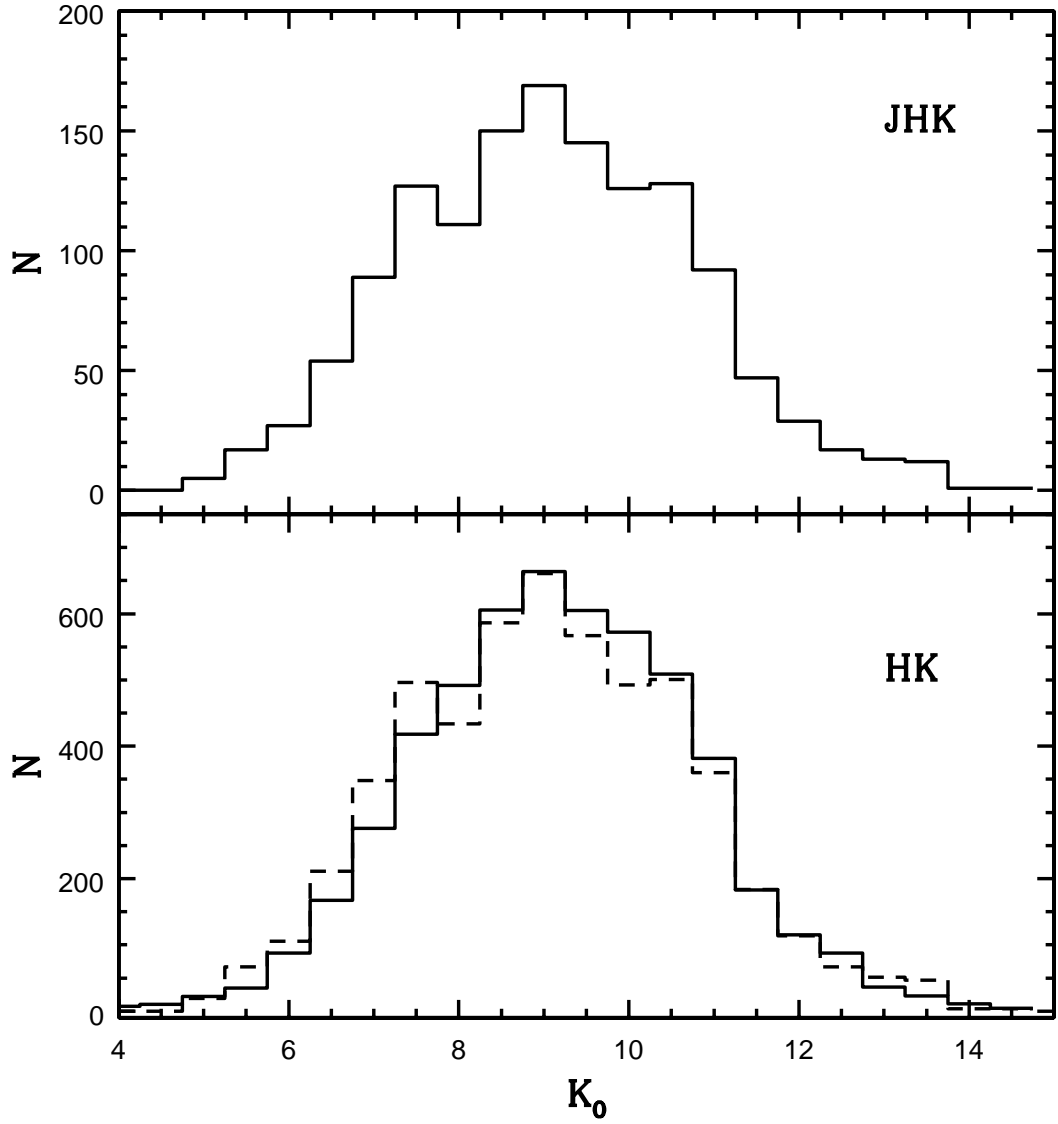} 
\caption{The reddening-corrected K$_o$ LFs, generated using the A$_K$ values 
shown in Figure 7. The dashed line in the lower panel shows the LF from the 
upper panel, scaled to match the number of stars having only $H-K$ colors. Note 
that the two LFs are very similar after accounting for differences in sample 
size.}
\end{figure}

\begin{figure}
\plotone{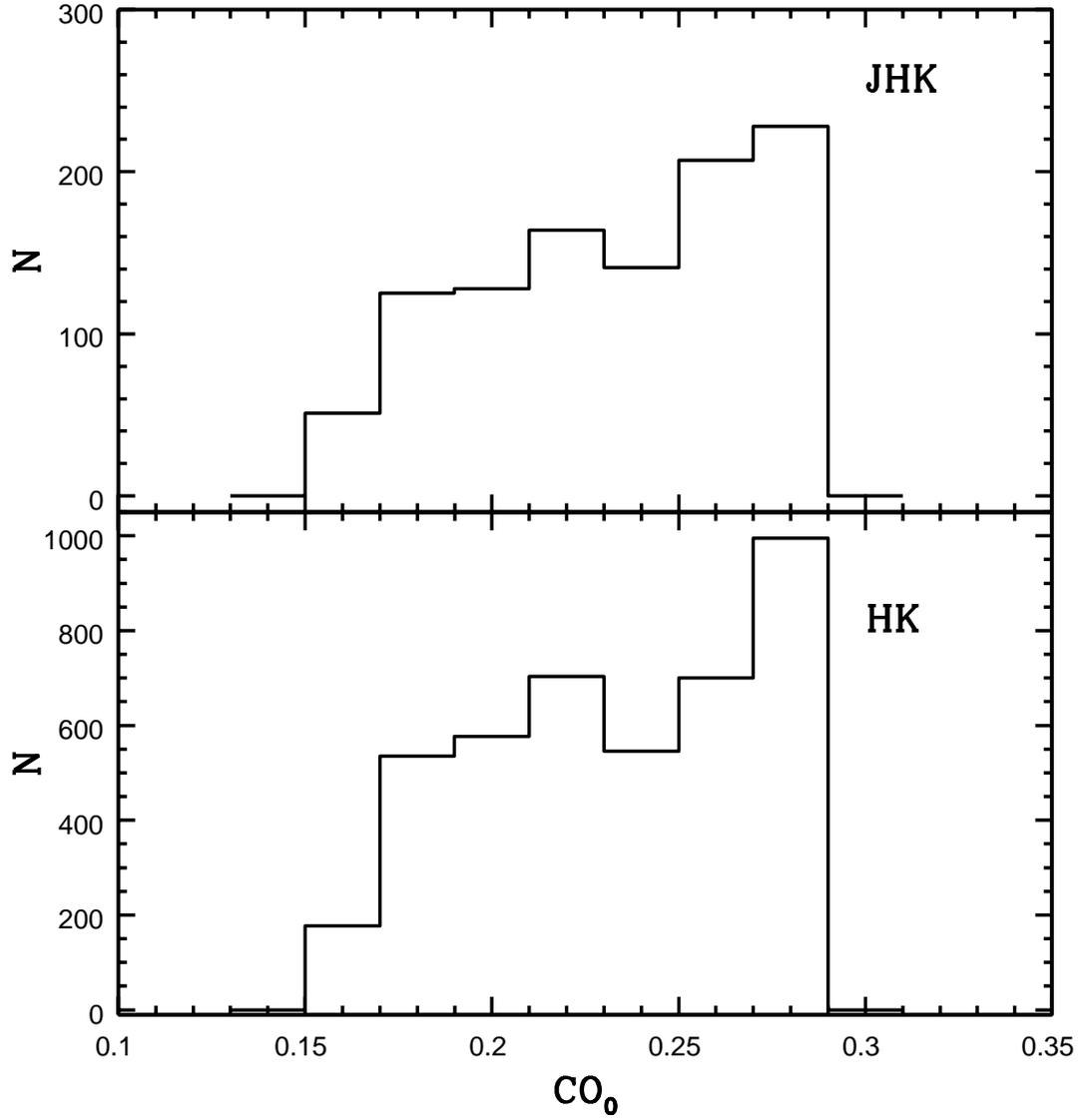} 
\caption{The CO$_o$ distributions predicted from the M$_K - CO$ relation 
given in Table 3B of Frogel \& Whitford (1987). These curves, which do not 
include the effects of observational errors in the CO indices, were constructed 
using the K$_o$ LFs shown in Figure 10 for stars with K$_o$ between 11.25 and 
7.25.}
\end{figure}

\begin{figure}
\plotone{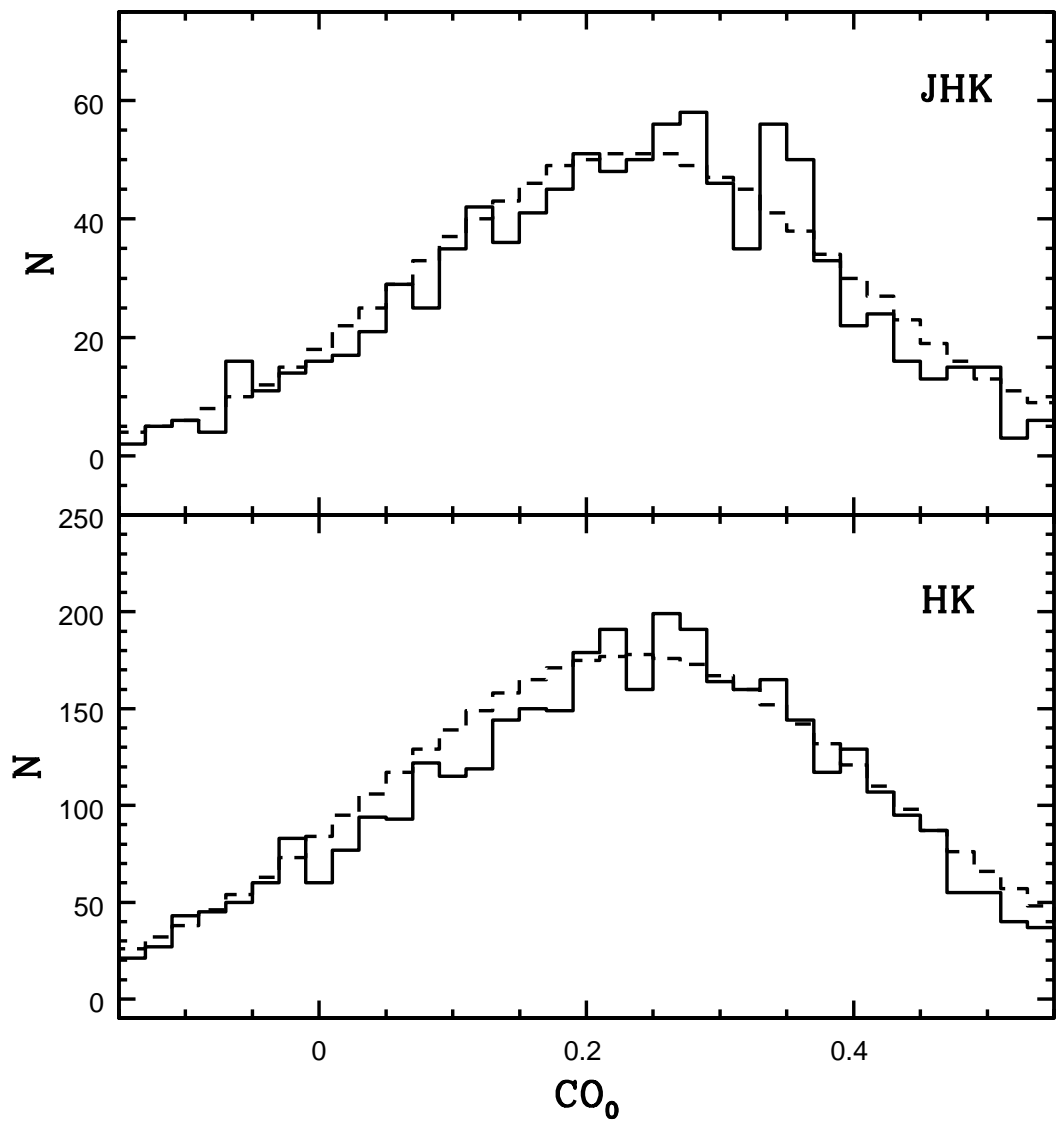} 
\caption{The CO$_o$ distribution for stars with $K_o$ 
between 11.25 and 7.25, which corresponds to the brightness range 
of M giants in BW. The top panel shows the CO$_o$ distribution 
if extinctions are computed using both $J-H$ and $H-K$ colors, while the lower 
panel shows the CO$_o$ distribution that results if only $H-K$ colors are used. 
The dashed line in each panel shows the predicted CO$_o$ distributions from 
Figure 11, smoothed to match the observed number of stars with CO$_o$ 
between 0.16 and 0.28.}
\end{figure}

\begin{figure}
\plotone{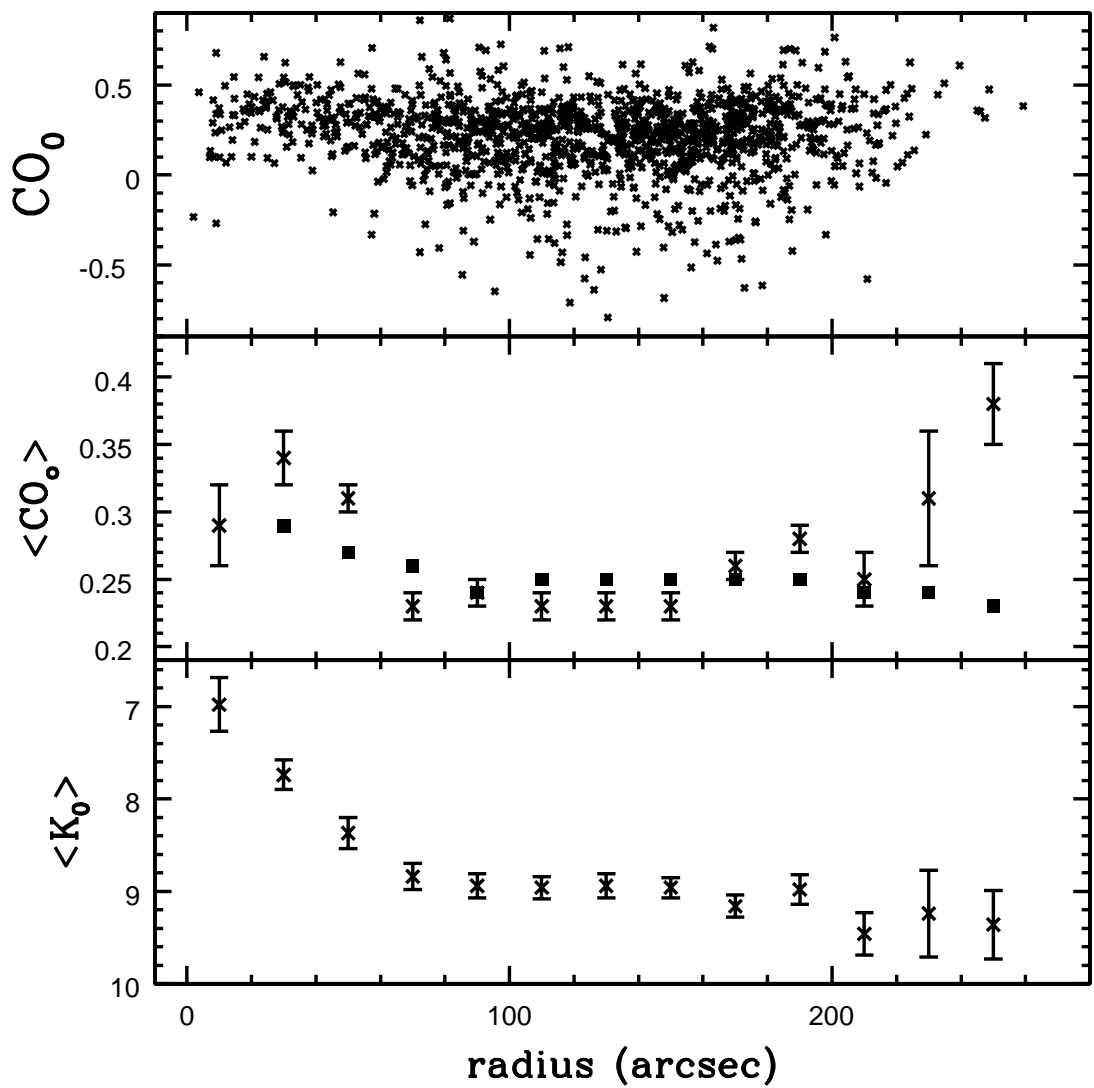} 
\caption{The top panel shows CO$_o$ as a function of distance from SgrA* for 
stars with extinctions estimated from both $J-H$ and $H-K$ colors. The mean 
values of CO$_o$ and $K_o$ in 20 arcsec intervals are plotted as crosses in the 
middle and lower panels, with the errorbars showing the uncertainty in the 
mean. The filled squares in the middle panel are the CO$_o$ values predicted 
from $<K_o>$, based on the M$_K - CO$ relation defined by M giants in BW.}
\end{figure}

\begin{figure}
\plotone{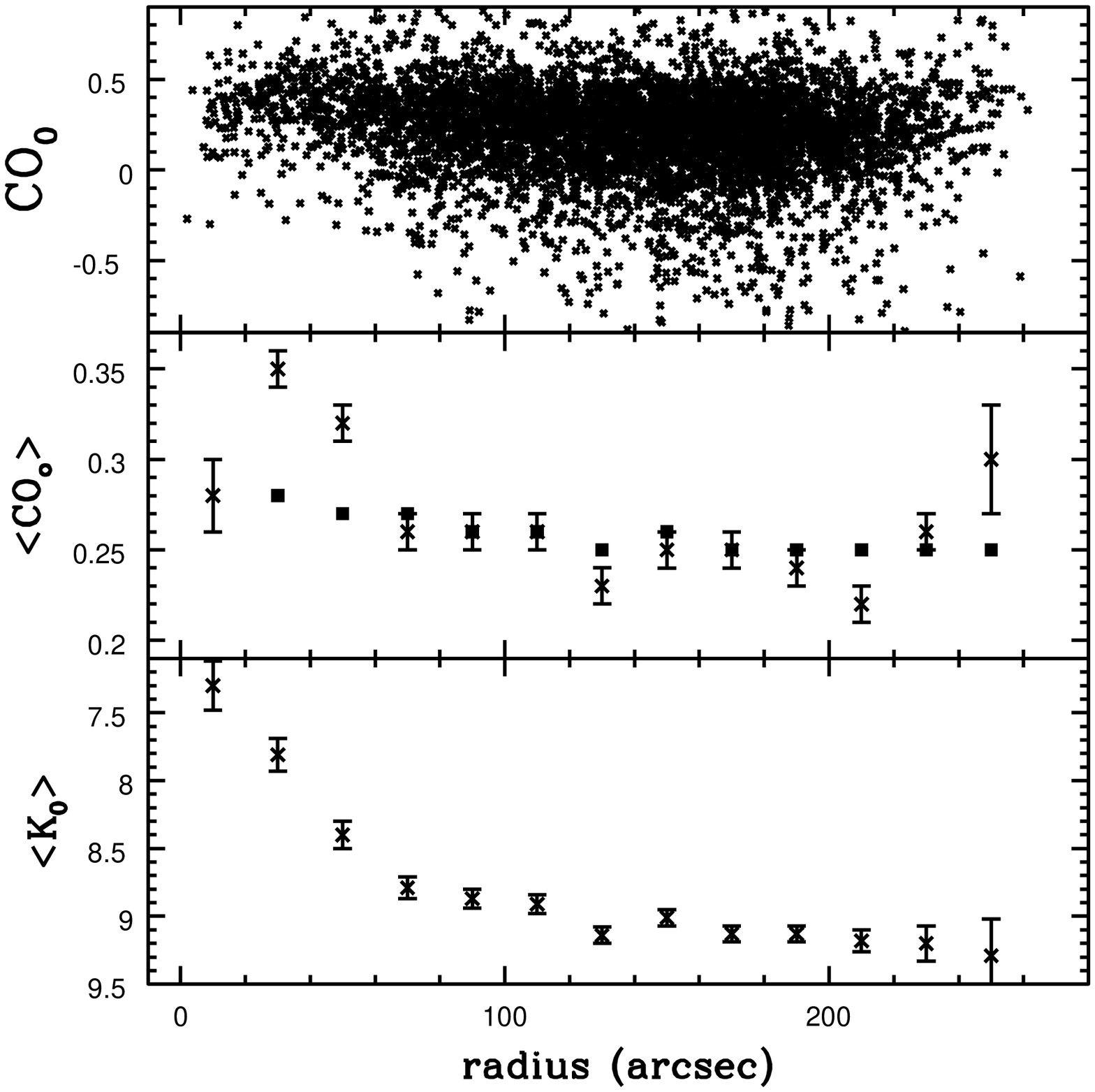} 
\caption{Same as Figure 13, except for stars that had extinctions estimated 
from $H-K$ colors only.}
\end{figure}

\clearpage
\begin{references}

\reference{} Allen, D. A. 1994, The Nuclei of Normal Galaxies: Lessons from the 
Galactic Center, R. Genzel \& A. I. Harris, 
Dordrecht:Reidel, 293

\reference{} Bergbusch, P. A., \& VandenBerg, D. A. 1992, \apjs, 81, 163

\reference{} Bertelli, G., Bressan, A., Chiosi, C., Fagotto, F., \& Nasi, E. 
1994, \aapr, 106, 275

\reference{} Blanco, V. M., \& Terndrup, D. M. 1990, \aj, 98, 843

\reference{} Blum, R. D., Sellgren, K., \& DePoy, D. L. 1996a, \apj, 470, 864

\reference{} Blum, R. D., Sellgren, K., \& DePoy, D. L. 1996b, \aj, 112, 1988

\reference{} Carr, J. S., Sellgren, K., \& Balachandran, S. C. 1996, The Galactic 
Center, ASP Conf. \# 102, R. Gredel, 212

\reference{} Casali, M., \& Hawarden, T. 1992, JCMT-UKIRT Newsletter, 4, 33

\reference{} Davidge, T. J. 1997, \aj, 113, 985

\reference{} Davidge, T. J., Simons, D. A., Rigaut, F., Doyon, R., \& Crampton, 
D. 1997, \aj , 114, 2586.

\reference{} DePoy, D. L., \& Sharpe, N. A. 1991, \aj, 101, 1324

\reference{} Eckart, A., Genzel, R., Hofmann, R., Sams, B. J., \& Tacconi-Garman, 
L. E. 1995, \apj, 407, L23

\reference{} Elias, J. H., Frogel, J. A., \& Humphreys, R. M. 1985, \apjs, 57, 91

\reference{} Elias, J. H., Frogel, J. A., Matthews, K., \& Neugebauer, G. 1982, 
\aj, 87, 1029

\reference{} Frogel, J. A., \& Whitford, A. E. 1987, \apj, 320, 199

\reference{} Frogel, J. A., Persson, S. E., \& Cohen, J. G. 1981, \apj, 246, 842

\reference{} Frogel, J. A., Persson, S. E., \& Cohen, J. G. 1983, \apjs, 53, 713 

\reference{} Frogel, J. A., Persson, S. E., Aaronson, M., \& Matthews, K. 1978, 
\apj, 220, 75

\reference{} Haller, J. W., Rieke, M. J., Rieke, G. H., Tamblyn, P., Close, L., 
\& Melia, F. 1996, \apj, 456, 194

\reference{} Hodapp, K.-W., Rayner, J., \& Irwin, E. 1992, \pasp, 104, 441

\reference{} Kraft, R. P. 1994, \pasp, 106, 553

\reference{} Minniti, D., Olszewski, E. W., Liebert, J., White, S. D. M., Hill, 
J. M., \& Irwin, M. J. 1995, \mnras, 277, 1293

\reference{} Morris, M., \& Serabyn, E. 1996, \araa, 34, 645

\reference{} Reid, M. 1993, \araa, 31, 345

\reference{} Rieke, G. H., \& Lebofsky, M. J. 1985, \apj, 288, 618
 
\reference{} Saha, P., Bicknell, G. V., \& McGregor, P. J. 1996, \apj, 467, 636

\reference{} Sellgren, K., McGinn, M. T., Becklin, E. E., \& Hall, D. N. B. 
1990, \apj, 359, 112

\reference{} Stetson, P. B. 1987, \pasp, 99, 191

\reference{} Stetson, P. B., \& Harris, W. E. 1988, \aj, 96, 909

\reference{} Tiede, G. P., Frogel, J. A., \& Whitford, A. E. 1995, \aj, 110, 2788

\reference{} van den Bergh, S. 1989, \araa , 1, 111

\end{references}
\end{document}